\documentclass[twocolumn]{article}

\usepackage{lmodern}
\usepackage{authblk}
\usepackage{amsmath}
\usepackage[colorlinks=true, allcolors=cyan]{hyperref}
\usepackage{graphicx,xcolor}
\usepackage{cite}
\usepackage[
    top=2.54cm,
    bottom=2.54cm,
    left=2.54cm,
    right=2.54cm
]{geometry}

\usepackage[normalem]{ulem}
\usepackage[utf8]{inputenc}
\usepackage{placeins}               
\usepackage[
    separate-uncertainty = true,
    locale               = UK,
    group-separator      = {,},
    exponent-product     = \cdot
]{siunitx}

\graphicspath{{./figures/}}

\title{\textbf{Efficient broadband second‑harmonic generation in a multi‑pass cell}}

\author[1,*]{Arthur Sch\"onberg}
\author[1]{Nikolas Rupp}
\author[1,2]{Ana Oliveira e Silva}
\author[1,2]{Lorenzo Pratolli}
\author[1]{\\Yujiao Jiang}
\author[2,3]{Vincent Wanie}
\author[2,3]{Terence Mullins}
\author[1,2,3]{Francesca Calegari}
\author[1,4,5]{\\Christoph M. Heyl}

\affil[1]{Deutsches Elektronen‑Synchrotron DESY, Notkestr. 85, 22607 Hamburg, Germany}
\affil[2]{Center for Free-Electron Laser Science CFEL, Deutsches Elektronen-Synchrotron DESY, Notkestr. 85, 22607 Hamburg, Germany}
\affil[3]{The Hamburg Centre for Ultrafast Imaging, Universität Hamburg, Luruper Chaussee 149, 22761 Hamburg, Germany}
\affil[4]{Helmholtz‑Institut Jena, Fr\"obelstieg 3, 07743 Jena, Germany}
\affil[5]{GSI Helmholtzzentrum für Schwerionenforschung GmbH, Planckstraße 1, 64291 Darmstadt, Germany}

\affil[*]{arthur.schoenberg@desy.de}   

\begin{document}
\twocolumn[
  \begin{@twocolumnfalse}
    \maketitle
    
\begin{abstract}
Second‑harmonic generation (SHG) is a widely used nonlinear optical process which has been optimized for high
frequency conversion efficiency, broad bandwidths as well as optimum spatial
and temporal quality of the generated light. 
However, limitations of the generation efficiency arise due to temporal dispersion and pulse walk‑off effects introduced by the nonlinear crystal. To circumvent these limitations, quasi‑waveguide schemes, in particular multi‑pass cells (MPCs), can be employed, providing flexible phase‑tuning capabilities. 
In this work, we demonstrate for the first time broadband and efficient SHG in an MPC. 
SHG efficiencies of \SI{72}{\%} (for \SI{63}{fs}, \SI{50}{\micro\joule} pulses) and \SI{47}{\%} (for \SI{15}{fs}, \SI{20}{\micro\joule} pulses) are obtained by simultaneously matching the relative phase and group‑delay between the fundamental and second‑harmonic fields over multiple passes through the MPC. 
Our work sets a new record for efficient Gaussian-beam, ultra‑short pulse SHG and provides a general route to address efficiency‑bandwidth limitations of nonlinear optical processes.
\end{abstract}
\vspace{5 mm}
\end{@twocolumnfalse}
  ]


\section{Introduction}
\begin{figure*}[t!]
    \centering
    \includegraphics[width=0.8\textwidth]{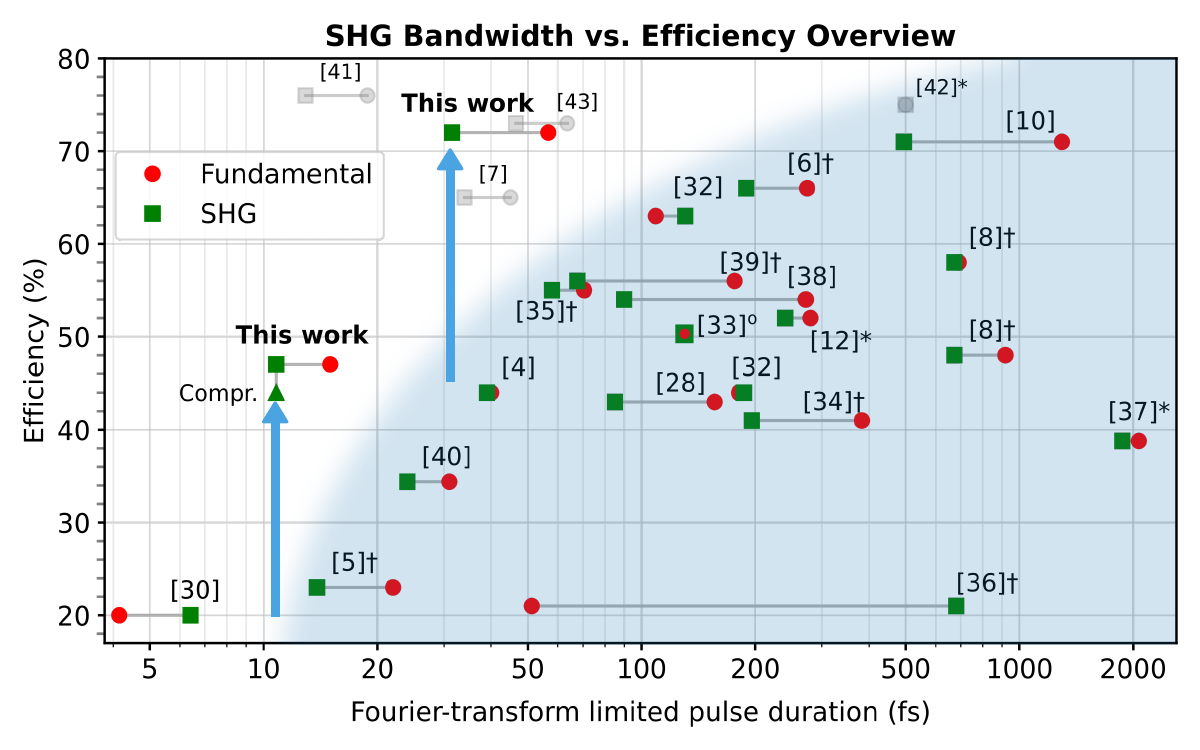}
    \caption{Overview of various works on second-harmonic generation (SHG), relating efficiency to the fundamental (red dots) input and SHG output (green squares) bandwidths, here expressed as the full-width half-maximum (FWHM) of the Fourier-transform limited (FTL) pulses. 
    The green triangle indicates the efficiency after compression of the SH pulse. The gray markers indicate flat-top or non-Gaussian beam systems. The blue arrows indicate the estimated efficiency improvement over the state of the art. $\dagger$The FTL was calculated using the spectral bandwidth provided in the reference. 
    $^*$The pulse duration was assumed as the FTL, since the reference did not provide the FTL or bandwidth.
    $^{\text{o}}$No FTL, bandwidth or pulse duration of the SH pulse provided.
    }
    \label{fig:overview}
\end{figure*}
Second‑harmonic generation (SHG) was first observed in 1961 \cite{Franken1961}, shortly after the invention of the laser \cite{Maiman1960}, and has since become a cornerstone of nonlinear optics.  
Today, SHG is indispensable for producing tunable mid‑infrared, visible and ultraviolet radiation \cite{Mason1994, Liu1997, Kanai2003, Rothhardt2011, Gobert2014, Gronloh2014, Ma2020, Roecker2020, Zhang2020, Karst2023, Xu2025}, for ultrafast pulse characterization \cite{Diels1985, Naganuma1989}, and for a variety of imaging and frequency‑comb applications \cite{Valev2012, Nuriya2016, Fortier2019, Ma2021}.  
Efficient $\chi^{(2)}$ conversion hinges on (i) a strong second‑order nonlinearity—typically provided by non‑centrosymmetric crystals—and (ii) fulfillment of the phase‑matching condition \cite{Giordmaine1962, Maker1962}.  
While many $\chi^{(2)}$ processes (SFG, DFG, optical rectification, OPA) share these requirements \cite{Boyd2020, Akhmanov1966}, achieving simultaneous phase and group‑delay (GD) matching in broadband, few‑cycle regimes remains challenging.  
Phase-matching can be achieved by various techniques, including type I, type II, noncritical birefringent phase-matching as well as quasi-phase-matching (QPM) \cite{Boyd2020}.
Free-space QPM has also been demonstrated using multi-pass cells \cite{Kovalenko2023,Pronin2024NonlinearOptical, Kovalenko2026}.
However, in the above examples, phase-matching can typically only be achieved for a certain spectral bandwidth, typically determined by the crystal length.
This leads to efficiency and bandwidth limits of the frequency-converted pulse \cite{Boyd2020}.
To overcome this challenge, in particular for SHG, methods such as adiabatic frequency conversion \cite{Leshem2016,Dahan2017} and prism-based phase-and group delay matching \cite{Zhang1990,Baum2004} have been developed. More recent works propose micro-structured fiberbased phase-and GD matching \cite{Tishchenko2022}.
While constituting promising approaches for efficient SHG, these methods either rely on single- or few-pass free-space geometries or wave-guides, where tunability of the relative phase between SHG and fundamental fields remains constrained, leaving the challenge to compromise between efficiency and bandwidth.

\nocite{Galletti2019,Ghotbi2004,Major2009,Zhu2004,Karst2023,Liu2020,Kanai2003,Dahan2017,Zhang2024,Gronloh2014,Roecker2020,Roecker2020a,Rothhardt2011,Liu1997,Zhao2022,Baum2004,Wang2025,Aparajit2021,Hillier2013,Mironov2011,Gobert2014} Here, we address this challenge by implementing SHG inside a Herriott-type multi-pass cell (MPC), enabling phase and group-delay compensation.
A $\chi^{(2)}$ nonlinear crystal is placed in the center of our second-harmonic multi-pass cell (SHG-MPC), enabling SHG for every pass through the MPC.
The SHG-MPC operates in a pressure-tunable gas cell filled with air or krypton. 
Group-delay (GD) compensating mirrors provide \SI{-15}{fs} relative GD per reflection between the fundamental and second-harmonic field. 
Fine tuning of phase and group delay is achieved by adjusting the gas pressure and crystal angle ($\theta$), enabling efficient, broadband SHG, ultimately limited by higher-order dispersion.

To demonstrate the SHG-MPC concept, we investigate two experimental regimes.
Using \SI{1030}{nm}, \SI{63}{fs}, \SI{50}{\micro J} fundamental pulses and a \SI{78}{\micro\meter} $\beta$-BaB$_2$O$_4$ (BBO) crystal, we achieve an SHG efficiency of 72\%. 
With shorter \SI{1030}{nm}, \SI{15}{fs}, \SI{20}{\micro J} driving pulses and a \SI{95}{\micro\meter} BBO crystal, we obtain 47\% SHG efficiency, with 44\% retained after compression.
In both cases, the generated second-harmonic output exhibits excellent pulse and beam quality. 
Our results demonstrate that MPCs are versatile tools for efficiently driving and tailoring SHG and potentially other second-order nonlinear processes.

Figure \ref{fig:overview} provides an overview over selected SHG works, comparing the bandwidth of the input and output pulses with the achieved conversion efficiencies.
For most works, a single-pass generation scheme is used. 
The trade-off between bandwidth and efficiency is clearly visible, highlighted by the blue background to guide the eye.
Compared with the works in Fig. \ref{fig:overview}, the SHG-MPC results obtained in this work show a notable increase in efficiency. For the two experimental regimes investigated here, the achieved efficiencies exceed the current state of the art by estimated factors of 1.6 and 2.2, respectively.

\begin{figure*}[t!]
    \centering
    \includegraphics[width=1.0\textwidth]{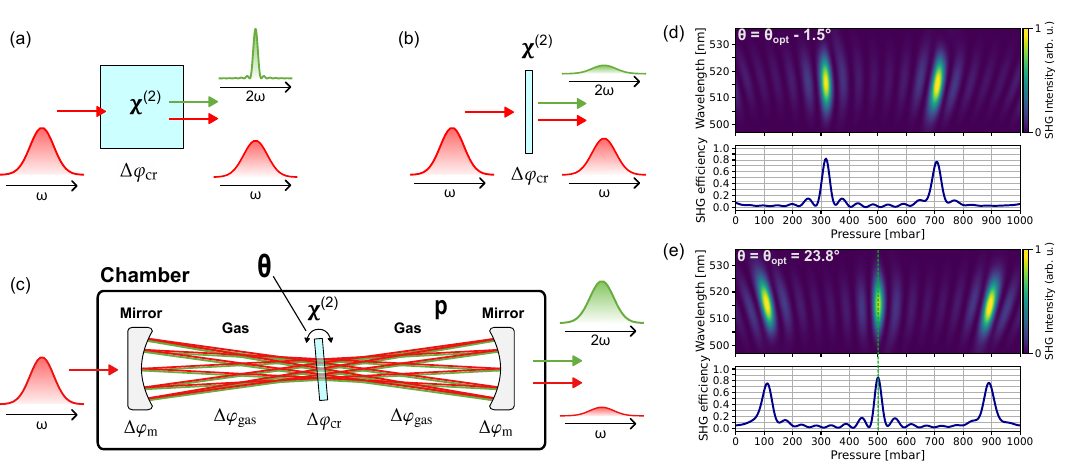}
    \caption{Illustration of the SHG-MPC concept and simulated spectrograms. (a) In a thick crystal, SHG can achieve high conversion efficiency, but the spectral bandwidth is limited by the narrow phase-matching bandwidth. (b) A thin crystal provides a much broader phase-matching bandwidth, but the reduced interaction length lowers the conversion efficiency. (c) Top view of the SHG-MPC, which combines high efficiency and broad bandwidth by appropriately compensating the phase and group-delay (GD) mismatch after each pass through a thin crystal.
    (d,e) Simulated idealized spectrograms of the spectrally resolved SH pulse as a function of gas pressure for two crystal angles. (d) Relative GD uncompensated. (e) Relative GD compensated. The simulation parameters match the experiment, except that ideal Gaussian beams are used as the fundamental input. The input pulse has a wavelength of \SI{1030}{nm}, a duration of \SI{30}{fs}, and an energy of \SI{25}{\micro\joule}. The MPC parameters are $R=\SI{1}{m}$ (radius of curvature), $N=4$ round trips ($2N=8$ passes) through a \SI{100}{\micro\meter}-thick BBO crystal, and a cavity length of $L=\SI{29.5}{cm}$. A relative GD of \SI{-15}{fs} is compensated at each mirror reflection, delaying the fundamental pulse. The simulations are performed using the model described in Supplementary Section S1. The optimal phase-matching angle is $\theta_{\mathrm{opt}}=23.8^{\circ}$.}
    \label{fig:concept}
\end{figure*}
\section{Concept: Second-harmonic-generation in multi-pass cells}
The SHG-MPC is characterized by its flexible phase- and amplitude tuning capabilities. 
This flexibility stems from the fact that the MPC acts as a quasi-waveguide in which the properties of the pulses can be adjusted at each pass though the cell using appropriate optical elements.
In particular, the cavity mirrors can be used to tune the phase and to compensate for the dispersion accumulated in the nonlinear crystal. In addition, two extra tuning parameters are introduced to precisely control the relative phase between the second-harmonic (SH) field and the fundamental field. 
These parameters are (i) the gas pressure $p$ in the vacuum chamber in which the SHG-MPC is set up and the critical angle $\theta$ relative to the optical axis of the (uniaxial) crystal.
Figure \ref{fig:concept} illustrates the SHG-MPC concept in comparison to single-pass SHG approaches. 
In Fig. \ref{fig:concept}(a), a long crystal is used, which narrows the spectral bandwidth of the SHG output but enables highly efficient SHG. 
In contrast, Fig. \ref{fig:concept}(b) shows that a thin crystal can convert the full bandwidth, but typically yields low conversion efficiencies due to the reduced interaction length. 
Fig. \ref{fig:concept}(c) illustrates the SHG-MPC in which both broad bandwidths and high conversion efficiencies can be achieved by tuning the the relative phases between the SH and fundamental fields.

In the SHG-MPC concept, we focus on the precise tuning of the relative phase (phase-matching) and the relative group-delay (GD-matching) between the SH and the fundamental fields.
In a nonlinear, birefringent crystal exhibiting a $\chi^{(2)}$-nonlinearity, phase-matching is typically achieved by propagating the fields in different axes with different dispersion relations $k_i(\omega)$.
Here, fundamental and second harmonic fields and the corresponding crystal axes are denoted with $i=1,2$, respectively.
We define the define the "zeroth"-order wavenumber-mismatch as
\begin{align}
\Delta k^{(0)} =\ 2k_1^{(0)}&(\bar\omega_1) - k_2^{(0)}(\bar\omega_2) = \nonumber \\[5pt]
=& \frac{\bar\omega_2}{c_0}\big(n_1(\bar\omega_1)-n_2(\bar\omega_2)\big)  \ ,
\label{eq:k0_mismatch}
\end{align}
where $\bar\omega_i$ is the central frequency of the respective field, $n_i$ the refractive index and $c_0$ the speed of light in vacuum.
For a given material length $L$, the zeroth-order phase-mismatch is $\Delta\phi^{(0)} = \Delta k^{(0)}\cdot L$.
In SHG, phase-matching is usually referred to the case when $\Delta k^{(0)} = 0$.
In this work, we focus on type I phase-matching, where the fundamental field $i=1$ is polarized orthogonally to the second-harmonic field $i=2$.
When phase matching is achieved, high conversion efficiencies are typically obtained for relatively narrowband pulses.
However, for ultrashort pulses, higher order phase-mismatch terms can become relevant, with the next order being the group-delay (GD) mismatch.
The GD wavenumber-mismatch between the fundamental and second-harmonic field can be written as
\begin{equation}
\Delta k^{(1)} = k_1^{(1)}(\bar\omega_1) - k_2^{(1)}(\bar\omega_2)  \ ,
\label{eq:k1_mismatch}
\end{equation}
where the accumulated GD in the medium is $\Delta\varphi^{(1)} = \Delta k^{(1)}\cdot L$.
In general, when $\Delta k^{(0)} = 0$, typically $\Delta k^{(1)} \neq 0$, i.e. when the phase is matched, the GD is not necessarily matched simultaneously.
In the temporal domain, a GD mismatch results in a temporal walk-off of the fundamental and second-harmonic pulses.
In the frequency domain, a GD mismatch appears as a linearly varying spectral phase mismatch, causing different frequency components of the pulses to walk off temporally during propagation. 
This leads to a narrowing of the acceptance bandwidth of the conversion process for long crystals.
Multi-pass cells provide the possibility of tuning the relative phase $\Delta \varphi$ of the two fields after each pass through the crystal, including the zeroth-order phase and the GD, overcoming the acceptance bandwidth limit compared to a single-pass generation scheme.
Assuming that the crystal is positioned in the center and both sides of the MPC are symmetric, the relative accumulated phase per pass amounts to
\begin{align}
    \Delta &\varphi_{\text{pass}}(\omega;\theta,p) = \nonumber \\[5pt] 
    &= \ \underbrace{\Delta \varphi_{\text{cr}}(\omega;\theta)}_{\text{Crystal}} \ + \ \underbrace{\Delta\varphi_{\text{gas}}(\omega;p)}_{\text{Gas}} \ + \ \underbrace{\Delta\varphi_{\text{m}}}_{\text{Mirror}}(\omega)
    \label{eq:deltaphi_mpc}
\end{align}
where we indicate the dependencies of each term on the two tuning parameters in the experiment, which are gas pressure $p$ and the "critical" crystal angle $\theta$ relative to the optical axis of the (uniaxial) crystal, as shown in Fig. \ref{fig:concept}(c).
The term $\Delta \varphi_{\text{cr}}(\omega;\theta)$ denotes the relative dispersion induced by propagation through the crystal depending on the angle $\theta$ of the wavevector $\vec{k}$ to the optical axis, $\Delta\varphi_{\text{gas}}(\omega;p)$ is the dispersion of the ambient gas depending on pressure $p$ and $\Delta\varphi_{\text{m}}(\omega)$ the phase induced by the mirror.
Note that we do not include the relative Gouy-phase contribution, as this is typically negligibly small within the interaction length in the crystal \cite{Schoenberg2026}.
\begin{figure*}[t!]
    \centering
    \includegraphics[width=0.78\textwidth]{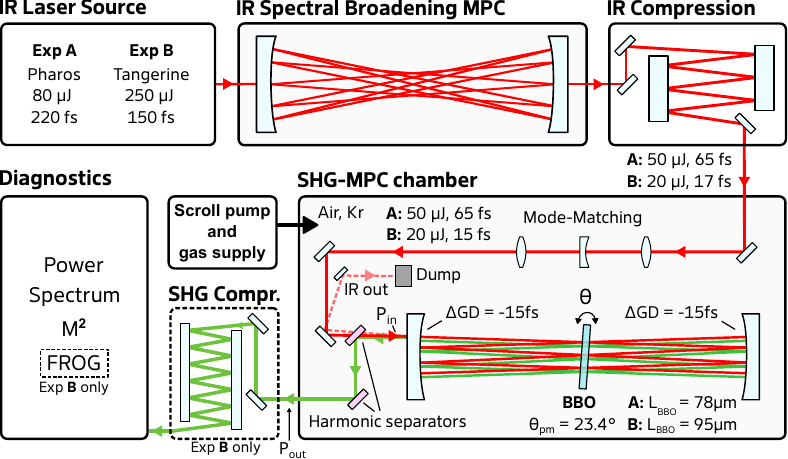}
    \caption{Experimental setup for both the long pulse (Exp A) and the short pulse (Exp B) experiments. Two different laser systems with similar parameters are used. After spectral broadening and post-compression, the pulses are sent into the chamber, where mode-matching and SHG in the SHG-MPC is carried out. Two harmonic separators are used to fully filter out the remaining fundamental light. In Exp B, the SH pulses are compressed using a chirped mirror compressor. The spectrum, M$^2$ and power is analyzed subsequently. Frequency-resolved optical gating (FROG) measurements are additionally carried out in Exp B.}
    \label{fig:setup65fs}
\end{figure*}

Equation (\ref{eq:deltaphi_mpc}) contains all contributions to the phase-mismatch between fundamental and second-harmonic field.
If we consider the zeroth and first order only, Eq. (\ref{eq:deltaphi_mpc}) can be expressed as the sum of the phase-mismatch $\Delta\varphi^{(0)} = 2\varphi_1^{(0)} - \varphi^{(0)}_2$ and the GD-mismatch $\Delta\varphi^{(1)} = \varphi_1^{(1)} - \varphi^{(1)}_2$ of each dispersive element in the SHG-MPC.
The phase‑ and group‑delay terms of the gas and crystal are obtained from their respective Sellmeier equations, while the contributions from the optics are taken from the coating specifications of the mirrors (see Supplement Fig. S6).
Phase-matching occurs when the following condition is fulfilled:

\begin{align}
    \Delta &\varphi^{(0)}_{\text{pass}}(\theta,p) = \nonumber \\[5pt] 
    &= \ \Delta \varphi_{\text{cr}}^{(0)}(\theta)\ + \ \Delta\varphi^{(0)}_{\text{gas}}(p) + \Delta\varphi^{(0)}_m \ \ \stackrel{!}{=} \ 2\pi n \ ,
    \label{eq:deltaphi3}
\end{align}

\noindent
where $n$ is an integer.
At the same time, for the first-order phase terms, the condition
\begin{align}
    \Delta &\varphi^{(1)}_{\text{pass}}(\theta,p) = \nonumber \\[5pt] 
    &= \ \Delta \varphi_{\text{cr}}^{(1)}(\theta)\ + \ \Delta\varphi^{(1)}_{\text{gas}}(p) + \Delta\varphi^{(1)}_m \ \ \stackrel{!}{=} \ 0
    \label{eq:deltaphi4}
\end{align}
needs to be fulfilled to ensure GD matching.
The amount of GD that is then required to be compensated by the mirrors can be determined by
\begin{align}
    \Delta\varphi^{(1)}_{\text{m}} = - \ \left(\Delta \varphi_{\text{cr}}^{(1)}(\theta_{\text{pm}})  \ + \ \Delta\varphi^{(1)}_{\text{gas}}(p_{\text{ref}})\right)\ ,
    \label{eq:gd_matching}
\end{align}
where $\theta_{\text{pm}}$ is the phase-matching angle ($23.4^{\circ}$ for BBO) and $p_{\text{ref}}$ the reference pressure of the gas that can be arbitrarily set.
In our experiments, we choose $p_{\text{ref}} = \SI{500}{mbar}$ of air, in order to enable a large phase tuning range above and below the GD matching point within the gas pressure tuning range supported by the setup of 0-1 bar.
A high conversion efficiency is then achieved by passing through the crystal multiple times.

Figures~\ref{fig:concept}(d,e) show simulated spectrograms and conversion efficiencies of the generated SH signal in a thin BBO crystal as a function of pressure $p$. 
In this example, the mirror phase is chosen such that the phase and GD conditions in Eqs.~(\ref{eq:deltaphi3}) and (\ref{eq:deltaphi4}) are simultaneously fulfilled at $p_{\text{opt}}=\SI{500}{mbar}$ and $\theta_{\text{opt}}=\SI{23.8}{}^{\circ}$.
Several main maxima of the SHG signal occur at pressures, where zeroth-order phase-matching condition defined in Eq.~(\ref{eq:deltaphi3}) is fulfilled. 
However, phase matching alone is not sufficient for efficient broadband SHG: the relative GD between the fundamental and SH pulses must also be compensated. 
If the crystal angle $\theta$ is not chosen appropriately, as illustrated in Fig.~\ref{fig:concept}(d), the GD condition of Eq.~(\ref{eq:deltaphi4}) is not fulfilled at the phase-matching pressure. 
In this case, pressure tuning can maximize the phase matching, but the fundamental and SH pulses remain temporally mismatched, limiting the bandwidth and conversion efficiency.
By adjusting the crystal angle to $\theta_{\text{opt}}$, the GD condition is shifted to coincide with the phase-matching condition at $p_{\text{opt}=\SI{500}{mbar}}$, as shown in Fig.~\ref{fig:concept}(e). 
At this point, both phase and GD matching are achieved simultaneously, resulting in broadband and efficient SHG.

Figures \ref{fig:concept}(d) and (e) furthermore illustrate the dispersion behavior of the SHG-MPC. The three dominant maxima in Fig. \ref{fig:concept}(e) are separated by approximately \SI{390}{mbar}, corresponding to a relative phase shift of $2\pi$ between the \SI{1030}{nm} fundamental and the \SI{515}{nm} SHG field over an MPC length of about \SI{29.5}{cm} in air. 
Below \SI{500}{mbar}, the spectra are tilted toward shorter wavelengths, indicating anomalous dispersion, whereas above this pressure the tilt reverses, corresponding to normal dispersion. 
Around \SI{500}{mbar}, where the group velocities are matched, both the conversion efficiency and the generated bandwidth reach their maximum values, indicating simultaneous phase and group-delay matching. 
Additional weaker maxima and minima arise between the main peaks. The number of these minor peaks is determined by the number of passes through the MPC.
A more detailed analytical treatment is provided in reference \cite{Schoenberg2026}.

\section{Experiment and Results}
For the experimental demonstration of the broadband SHG-MPC we conduct two experiments with two different pulse durations and bandwidths at \SI{1030}{nm} wavelength.
In the first experiment (referred to as experiment A), we perform SHG of $\SI{50}{\micro J}$, \SI{63}{fs} pulses with a Fourier-transform limit (FTL) of about \SI{56}{fs}.
Here we reach a conversion efficiency of \SI{72}{}\%.
In the second experiment (referred to as experiment B), we perform SHG using \SI{15.4}{fs} pulses, yielding an FTL of \SI{10}{fs} with an efficiency of \SI{47}{}\%. In experiment B we additionally compress the pulses using chirped mirrors down to \SI{13.8}{fs}.
In both experiments, the fundamental pulses are first post-compressed using a self-phase modulation (SPM) -based MPC post-compression system.
In experiment B, the post-compressed laser system developed in reference \cite{Silletti2023} is used.
For the experimental validation of broadband SHG‑MPC we investigate two distinct parameter sets at a wavelength of \SI{1030}{nm}.  
Experiment A employs \SI{50}{\micro\joule}, \SI{63}{fs} fundamental pulses (Fourier‑transform limit \SI{56}{fs}) as the input to the SHG-MPC.  
Experiment B uses \SI{20}{\micro\joule}, \SI{15.4}{fs} fundamental pulses (Fourier‑transform limit \SI{14.5}{fs}), with the generated second‑harmonic light subsequently compressed. 
In both configurations the fundamental pulses are first post-compressed in a conventional self‑phase‑modulation (SPM) multi‑pass cell (MPC) \cite{Silletti2023}.

Figure \ref{fig:setup65fs} displays the common experimental layout for both configurations. After spectral broadening and compression, the fundamental pulses are directed into the SHG‑MPC chamber, that can be filled with the selected gas at pressures ranging from 0 to \SI{1}{bar}. 
Inside this chamber, the beam is mode-matched and guided to the SHG-MPC.
Before coupling into the SHG-MPC, the beam passes through a first harmonic separator (HS).
The SHG-MPC consists of two concave mirrors with a radius-of-curvature (ROC) of $R=\SI{1}{m}$ and is configured to $N=4$ round-trips (8 passes) with a configuration parameter of $k=1$ \cite{Viotti2022}.
This corresponds to an MPC length of about $L=\SI{29.5}{cm}$.
The mirror coatings are specifically designed to compensate a group-delay (GD) of about \SI{-15}{fs} between \SI{1030}{nm} and \SI{515}{nm}, which is the relative GD accumulated over a single pass through the MPC including dispersion in the SHG crystal at a reference pressure of \SI{500}{mbar} of air.
As nonlinear crystals two Beta-Barium Borate ($\beta\text{-BaB}_2\text{O}_4$, BBO) crystals with thicknesses of $L_{\text{cr}}=\SI{78}{\micro m}$ and $L_{\text{cr}}=\SI{95}{\micro m}$ are used, both cut to the type I phase-matching angle of $\theta_{\text{pm}} = 23.4^{\circ}$ used without substrate.
The thinner BBO crystal is employed in experiment A, whereas the thicker crystal is used in experiment B, as this configuration experimentally yielded the optimum results.
The crystal is placed in the center of the SHG-MPC.
In this MPC configuration, the beam is weakly focused and the $1/e^2$ beam sizes are rather small with $\SI{385}{\micro m}$ beam radius at the MPC mirrors and $\SI{340}{\micro m}$ waist radius in the focus.
The BBO crystal is mounted such that the incidence angle $\theta$ can be adjusted, i.e. $\theta = \theta_{\text{pm}} + \Delta\theta$, where $\Delta\theta$ is the detuning from the phase-matching angle.
After \SI{8}{} passes through the SHG-MPC and the crystal, both fundamental and SHG beams are coupled out. 
The fundamental field is filtered using two harmonic separators in order to obtain a pure SHG beam.

We conduct measurements using both air (Experiment A and B) and Krypton (A).
In order to record spectrograms as, for example, in Fig. \ref{fig:concept}(e), we scan the gas pressure of the chamber from about \SI{50}{mbar} to \SI{1}{bar}. In these spectrograms, the SHG signal appears as multiple spectrally dependent maxima. 
While a signal maximum at a fixed wavelength reveals phase-matching between fundamental and SHG fields, the spectral tilt of these maxima is a direct indicator for GD matching. By tuning the crystal angle $\theta$ such that the spectrum in one of the maxima appears vertical, GD matching can be achieved.
In order to match both phase- and GD, we tune the pressure to the vertical maximum.
Afterwards, the SHG beam is characterized using various beam and pulse diagnostics.
In experiment B, we further compress the SHG pulse using a pair of chirped mirrors with an average group-delay dispersion (GDD) of \SI{-40}{fs^2} per reflection.

\subsection*{A. Long Pulse Conversion}
\label{sec:expA}
In Experiment A we use post-compressed pulses with \SI{63}{fs} pulse duration and an FTL of \SI{56}{fs}.
We attenuate the pulse energy to $\SI{50}{\micro J}$.
The repetition rate is set to \SI{4}{kHz}.
Here, we use the short BBO with $L_{\text{cr}}=\SI{78}{\micro m}$ thickness.
By recording the spectrum and the output power simultaneously and scanning the pressure in the chamber, we obtain similar spectrograms as shown in Fig. \ref{fig:concept}(e).
We use both air and Krypton as gas media in the SHG-MPC chamber.
While air is a more practical gas to use, Krypton provides larger dispersion, resulting in four phase-matching maxima within our fixed pressure tuning range. 
\begin{figure}[tb!]
    \centering
    \includegraphics[width=0.47\textwidth]{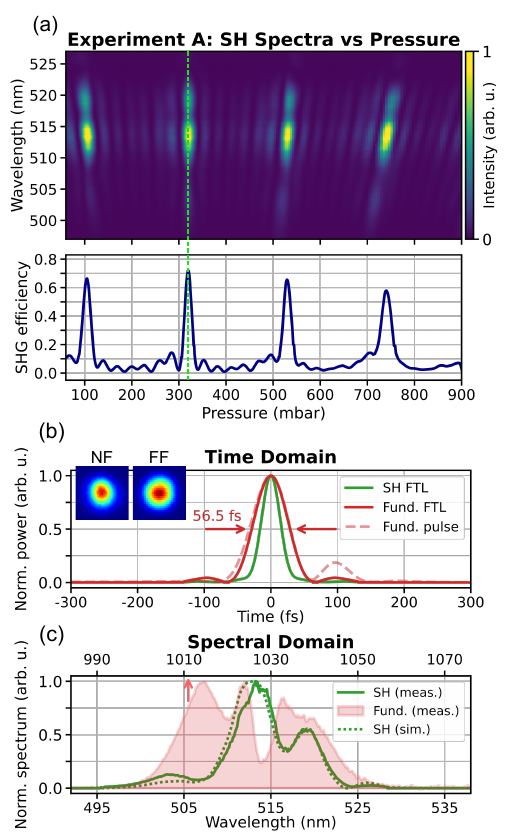}
    \caption{Experimental results of experiment A. (a) Spectrograms containing the spectrally-resolved SHG signal recorded as a function of gas pressure (here: Krypton) and the corresponding measured efficiencies displayed in the line-out plot. The green dashed line indicates the highest efficiency reached  after GD matching at around \SI{315}{mbar}. (b) Measured input and output pulses in the temporal domain. In this experiment the SH pulses are not compressed, thus the FTL pulses are shown. The insets show the beam profile of the SH beam in the near-field (NF) and far-field (FF). (c) Input and output spectrum. The simulated spectrum is in very good agreement with the measurement.}
    \label{fig:figure60fs}
\end{figure}

Figure \ref{fig:figure60fs} shows the experimental results for Krypton. 
In Fig. \ref{fig:figure60fs}(a), the spectrogram is shown together with the corresponding measured SHG efficiency.
In Fig. \ref{fig:figure60fs}(b-c) the fundamental and SH pulses at the largest efficiency are plotted in temporal and spectral domain.
Since the SH pulses are not compressed here, the FTL pulse is plotted along with the measured input fundamental pulse.
The bandwidth of the resulting SH pulses around \SI{515}{nm} support an FTL of \SI{31.5}{fs}. 
We measure a maximum conversion efficiency of \SI{72}{}\% at a pressure of $p=\SI{315}{mbar}$ in krypton.
The efficiency $\eta = P_{\text{in}}/P_{\text{out}}$ is determined using the input power $P_{\text{in}}$ at the position right before coupling into the SHG-MPC and the output power $P_{\text{out}}$ right after the vacuum chamber, both indicated in Fig. \ref{fig:setup65fs}.
We measure the beam quality of the SH beam using a commercial $M^2$ device (DataRay WinCamD-IR-BB) after out-coupling of the vacuum chamber revealing an excellent beam quality of $M^2<1.1$ of the output SH beam.
The simulated spectrum in Fig. \ref{fig:figure60fs}, as well as the simulated spectrograms (Fig. S1 in the supplement) are in excellent agreement.
Here we use the measured fundamental input pulses as the input and simulate the experimental conditions, using the (3+1)D nonlinear pulse propagation model described in the supplement section S1.
The resulting FTL of the simulated pulse is \SI{31.5}{fs} in agreement with the experimentally obtained spectrum.
We further simulate the same experiment excluding effects of SPM in order to investigate its influence on the generation process.
Here, the simulations show that the maximum efficiency does not change, while the output FTL duration increases to \SI{37}{fs}.
We therefore conclude that the nonlinear dynamics in this experiment are dominated by SHG, whereas SPM has a secondary influence, decreasing the Fourier-transform-limited pulse duration by about 15\%.

\subsection*{B. Short Pulse Conversion}
\begin{figure}[tb!]
    \centering
    \includegraphics[width=0.46\textwidth]{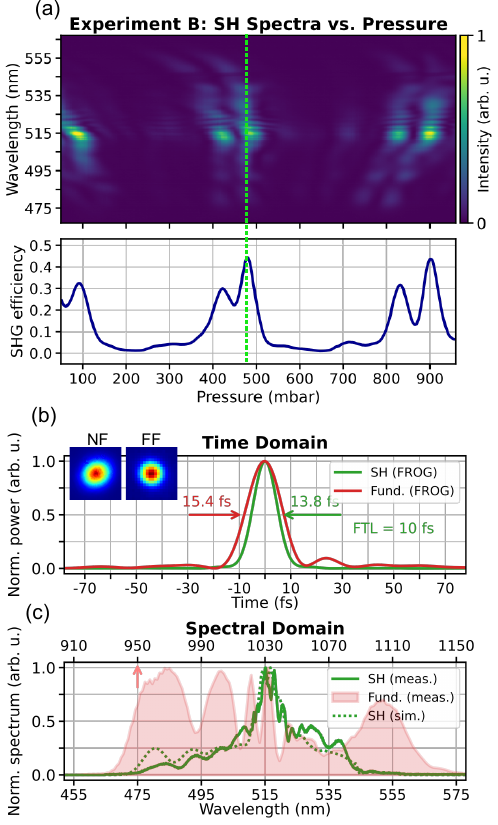}
    \caption{Experimental results of experiment B. (a) Spectrograms recorded as a function of gas pressure (here: air) and the corresponding measured efficiencies. The green dashed line indicates the highest efficiency at around 490 mbar. (b) Measured input and output pulses in the temporal domain. The insets show the beam profile of the SH beam in the near-field (NF) and far-field (FF). (c) Measured input and SH output spectra, as well as the simulated SH spectrum. The simulated spectrum is in excellent agreement with the experiment.}
    \label{fig:figure15fs}
\end{figure}

In Experiment B we use post-compressed pulses with \SI{15.4}{fs} pulse duration, an FTL of \SI{14.5}{fs} and a pulse energy of $\SI{20}{\micro J}$.
The post-compressed laser system was described earlier by Siletti et al. \cite{Silletti2023}.
The repetition rate of the system is set to \SI{5}{kHz}.
Here, we use air as the dispersive medium in the SHG-MPC, resulting in fewer phase-matched maxima due to the lower dispersion of air compared to Krypton.
We further change the crystal to the longer one with $L_{\text{cr}}=\SI{95}{\micro m}$.
The experimental results are shown in Fig. \ref{fig:figure15fs}.
In the recorded spectrograms (Fig. \ref{fig:figure15fs}(a)), we observe a global maximum at around \SI{490}{mbar}.
The signal maxima  appear curved, effectively reducing the bandwidth and efficiency of the SHG process at a fixed gas pressure. 
The curvature results from higher-order relative phase terms between fundamental and SH fields, with the main contribution arising from a relative group-delay dispersion (GDD).
The GDD (and higher phase orders) result from both, the phase characteristics of the mirror coatings and the dispersion of the materials in the MPC, where the latter is responsible for the major part of the GDD mismatch. While the MPC mirrors are designed for GD matching, higher order phase terms are not compensated.
The spectra in Fig. \ref{fig:figure15fs} exhibit modulations that can be attributed to both the modulation of the post-compressed input spectrum as well as the phase introduced by the mirrors.
The efficiency maxima exhibit a double-peak structure, which we are able to reproduce in our simulations by implementing the angle variation of the critical $\theta$ angle, introduced by the MPC beam geometry (supplement Fig. S3).
According to our simulations, this effects does not significantly reduce the SHG conversion efficiency, but can introduce a second peak as seen in the experimental spectrograms as seen in Fig. \ref{fig:figure15fs}(a).
In section \ref{sec:discussion}, we discuss how to eliminate the angular variation of the critical angle $\theta$ in the SHG-MPC.
\begin{figure*}[t!]
    \centering
    \includegraphics[width=\textwidth]{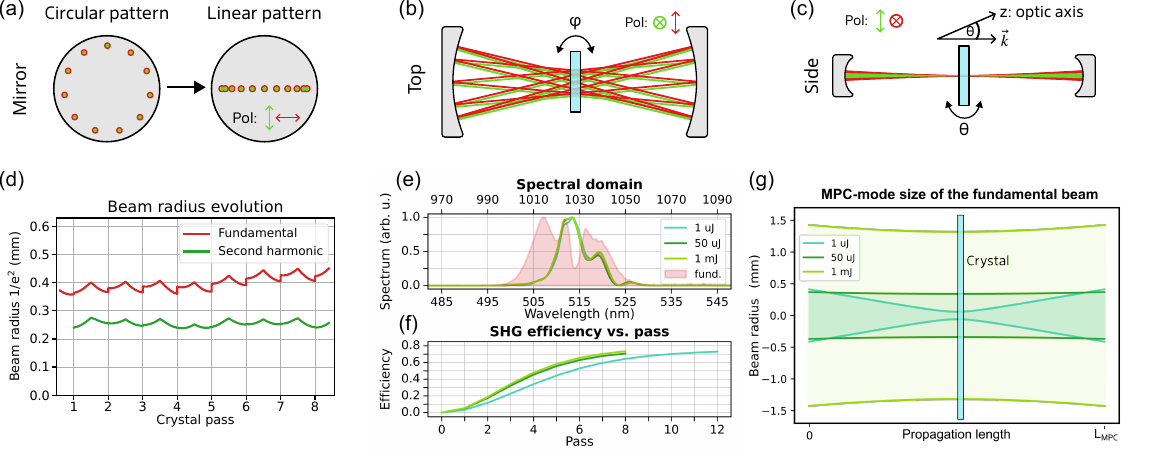}
    \caption{Optimized geometry and corresponding beam properties for an SHG-MPC employing a uniaxial crystal (e.g., BBO). The polarizations of the fundamental (red) and SH (green) beams are indicated by arrows, with the SH polarized along the extraordinary axis. Crosses denote polarization vectors pointing into the plane of the figure. (a) Comparison of a conventional round beam pattern, typically used in standard MPC configurations, with the proposed linear pattern. (b) Top view of the MPC with the linear pattern. From this perspective, the linear and round patterns appear identical. (c) Side view of the MPC, where the linear pattern appears like a single beam at a fixed angle $\theta$. (d) Simulated beam-radius evolution of the fundamental and SH beams throughout the SHG-MPC using the parameters of Experiment A.
    (e–g) Simulations of energy-scaled SHG-MPCs using the measured fundamental spectrum from Fig.~\ref{sec:expA} as the simulation input. (e) Simulated MPC output spectra for three input pulse energies (\SI{1}{\micro\joule}, \SI{50}{\micro\joule}, and \SI{1}{\milli\joule}), together with the input spectrum. (f) Corresponding evolution of the SHG conversion efficiency per pass. (g) Comparison of the fundamental (\SI{1030}{\nano\meter}) MPC beam radii for the three pulse energies. The MPC geometries are chosen such that the maximum peak intensity in the crystal is approximately \SI{600}{\giga\watt\per\square\centi\meter}, while the crystal thickness is fixed at \SI{100}{\micro\meter} in all cases.}
    \label{fig:linear_pattern}
\end{figure*}
After setting the gas pressure to the efficiency maximum (indicated in Fig. \ref{fig:figure15fs}(a)), we compress the pulse using a pair of chirped mirrors with an average GDD of \SI{-40}{fs^2} per reflection and a total of 12 reflections, yielding about \SI{-480}{fs^2} of compensated GDD.
We characterize the SH pulses using frequency-resolved optical gating (FROG) device, where we retrieve a pulse duration of $\SI{13.8}{fs}$ and a conversion efficiency of \SI{47}{\%}.
This efficiency is measured before the chirped mirror compressor. 
The total compressor transmission amounts to 93.4\% , which can still be optimized by employing higher reflectivity chirped mirrors.
The bandwidth measured with a spectrometer supports a pulse duration of \SI{10}{fs}, while the retrieved spectrum from the FROG measurement reveals an FTL of \SI{12.6}{fs}.
We attribute this discrepancy to the limited phase-matching bandwidth of the SHG crystal in the FROG setup. The simulated FTL duration of \SI{10}{fs}, in agreement with the spectrometer-based estimate, further supports this interpretation.
When SPM is suppressed in the simulations, the resulting FTL duration increases to \SI{11.5}{fs}, showing again that SPM only plays a minor role in the SHG MPC.
We measure a very good beam quality of $M^2 = 1.18\pm 0.12$ for the SH output, compared to about $M^2 = 1.16$ for the fundamental input beam.

\section{Discussion and Outlook}
\label{sec:discussion}
The SHG-MPC enables efficient nonlinear frequency-doubling of broadband pulses by exploiting the phase-tuning capabilities of MPCs.
Instead of passing through a relatively thick nonlinear medium once, the propagation is split into many passes, where in-between each pass, the free propagation of the pulses and reflection at the MPC mirror is used to compensate the accumulated phase in the nonlinear crystal.
In this way, the SHG-MPC enables to overcome the efficiency-bandwidth limit of single-pass SHG schemes.

A challenge that may arise is due to the typically round beam pattern geometry of the MPC, which can lead to considerable angles variations at the SHG crystal, changing the propagation axis of the beams and thus the dispersion relations in a non-isotropic crystal.
This issue, however, can be circumvented by two measures. 
First, the beam geometry can be changed to a linear pattern, as shown in Fig. \ref{fig:linear_pattern}(a)-(c). 
Second, uniaxial crystals -- for example BBO -- can be used, with the beam pattern aligned such that the angles vary only along the $\phi$-axis, while the $\theta$-axis, along which the extraordinary SH field propagates, remains constant for each pass.
The angle $\theta$ again refers to the angle of the $\Vec{k}$-vector to the optical axis $z$ of the crystal and $\phi$ is the angle relative to the $xz$ plane \cite{Saleh2019,Maldonado1991}.
In this way, extraordinary propagation of the SH field takes place at a defined $\theta$-angle, leading to unchanged dispersion properties in the crystal at each pass.
In the experiments, the MPC configuration is chosen such that the angular variation is small ($<\pm 0.5^{\circ}$) by making the beam pattern on the mirrors as compact as possible, while still maintaining efficient in/out-coupling.
While a double-peak in the efficiency (Fig. \ref{fig:figure15fs}) appears -- due to the remaining angle variations in the $\theta$-direction -- the efficiency does not significantly reduce.
However, for different configurations, where the angles can be large, e.g. in tight-focusing geometries, the linear pattern approach should be implemented.

Another interesting aspect to discuss is mode-matching of the second harmonic in the SHG-MPC.
In general, the input beam sent into an MPC needs to be matched to the transversal mode defined by the cell.
In the context of the SHG-MPC, the question arises whether the generated SH field satisfies the mode-matching conditions as well.
In our (3+1)D simulations, taking into account all relevant spatio-temporal pulse propagation effects (see supplementary section S1), we show that the SH automatically fulfills the mode-matching condition, if the fundamental beam is mode-matched to the MPC.
Figure \ref{fig:linear_pattern}(d) shows the simulated beam radius over all eight passes in the SHG-MPC.
Here we simulate the scenario of experiment A (section \ref{sec:expA}A) and show that the SH mode follows the fundamental mode and is very well mode-matched.
As expected for SHG in the low conversion limit, the mode radius of the SH beam is reduced by a factor $\sqrt{2}$ relative to the fundamental mode, enabling to satisfy mode-matching for the SHG field in the nearly linear regime, where Kerr lensing is negligible \cite{Viotti2022}.
The abrupt increases of the radius of the fundamental beam observed in Fig. \ref{fig:linear_pattern}(d) arise from the conversion process. 
As SHG depletes the central portion of the fundamental beam, the remaining intensity distribution broadens, resulting in an increased effective beam radius.
\begin{figure}[tb!]
    \centering
    \includegraphics[width=0.45\textwidth]{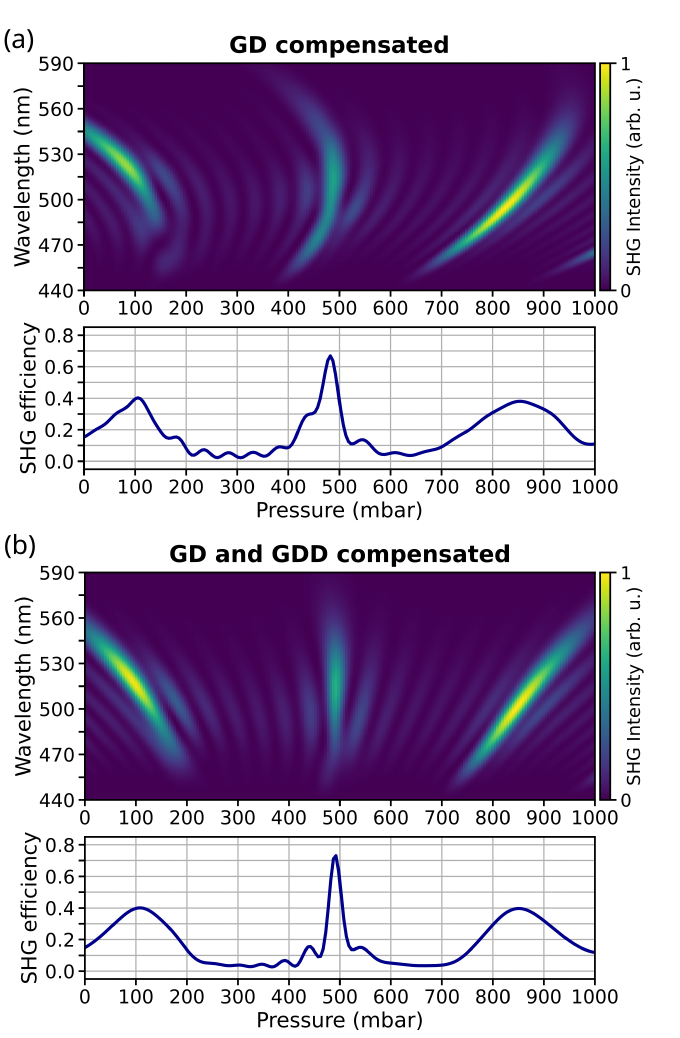}
    \caption{Simulated spectrogram traces and efficiencies for a \SI{7}{fs} Gaussian pulse input using a $\SI{100}{\micro m}$ BBO crystal and air as dispersive gas. (a) GD compensated with remaining GDD. (b) Both GD and GDD compensated using formulas Eq. (\ref{eq:gd_matching}) and Eq. (\ref{eq:gdd_matching}).}
    \label{fig:spectrograms_idealized_7fs}
\end{figure}
Another relevant aspect to explore is energy-scaling of the SHG-MPC.
In our experiments, we use pulses with moderate pulse energies of a few 10s of $\SI{}{\micro J}$.
However, the pulse energy is not limited to this range, but can be scaled to either lower or higher pulse energies.
For lower (higher) pulse energies, the MPC can be configured to tighter (looser) focusing geometries in order to reach suitable intensities inside the crystal. 
Energy up-scaling can be done in a compact way by using convex-concave (cx-cv) MPCs \cite{Hariton2023} or by scaling up the radius-of-curvature (ROC) and the MPC length, where the mode-size scales as a square-root with the ROC \cite{Heyl2022}, optionally combined with beam path folding in order to decrease the MPC length \cite{Schoenberg2025}.
The important parameter for scaling is the peak intensity of the fundamental field in the nonlinear crystal, which we try to keep around $\SI{600}{GW/cm^2}$.
Fig. \ref{fig:linear_pattern}(e) displays simulated MPC output spectra for low ($\SI{1}{\micro J}$), moderate ($\SI{50}{\micro J}$) and high energy (\SI{1}{mJ}), revealing very similar characteristics.  
For the \SI{50}{\micro J} case, we consider the same SHG-MPC configuration and input pulse as used in experiment A. 
The high energy case uses an up-scaled version of this configuration by increasing the mirror ROC to $R=\SI{4.5}{m}$ and an MPC length to about \SI{1.3}{m}, where we keep the number of round-trips at $N=4$ and $k=1$.
In the low energy case, we consider $R=\SI{50}{mm}$, $N=6$ round-trips and $k=5$, resulting in an MPC length of about \SI{10}{cm} and strongly focusing geometry.
Figure \ref{fig:linear_pattern}(f) shows the efficiency per pass.
The number of passes is chosen to reach saturation of the efficiency.
Figure \ref{fig:linear_pattern}(g) displays the eigenmode structure of each MPC in the focal region.
We calculate efficiencies of > 70\% for all three configurations, using the same experimentally measured input temporal profiles as in our experiment (experiment A, section \ref{sec:expA}) with \SI{63}{fs} input pulse duration and an FTL of \SI{56}{fs}.

In addition to energy up-scaling, an interesting development direction for the SHG-MPC is pulse duration down-scaling.
The main challenge arises from higher-order phase-mismatch introduced by the dispersion of the crystal, gas and the MPC mirrors.
For very short pulses, higher order relative phases terms such as the second-order phase (group-delay-dispersion, GDD) and third order (third-order-dispersion, TOD) can limit the bandwidth and the SHG conversion efficiency. 
Similar to GD, the GDD can be matched approximately by setting the second-order mirror dispersion to
\begin{align}
    \Delta\varphi^{(2)}_{\text{m}} \approx - \ \left(\Delta \varphi_{\text{cr}}^{(2)}(\theta_{\text{pm}})  \ + \ \Delta\varphi^{(2)}_{\text{gas}}(p_{\text{ref}})\right)\ ,
    \label{eq:gdd_matching}
\end{align}
where $\Delta \varphi_{\text{cr}}^{(2)}(\theta_{\text{pm}})$ and $\Delta\varphi^{(2)}_{\text{gas}}(p_{\text{ref}})$ is the accumulated GDD in the crystal and the gas respectively, at phase-matching angle $\theta_{\text{pm}}$ and reference pressure $p_{\text{ref}}$.
Figure \ref{fig:spectrograms_idealized_7fs} shows simulated spectrograms  for a two-cycle, \SI{7}{fs} input pulse centered at \SI{1030}{nm} wavelength. 
In this simulation, the MPC configuration and dispersion properties are similar to experiment B, however, here we use an idealized mirror dispersion, which compensates the relative GD by \SI{-15}{fs} (Fig. \ref{fig:spectrograms_idealized_7fs}(a) and (b)) and in addition the relative GDD by about \SI{-12}{fs^2} in \ref{fig:spectrograms_idealized_7fs}(b).
In Fig. \ref{fig:spectrograms_idealized_7fs}(a) the higher-order phase-mismatch can be identified by the curvature of the signal maxima in the spectrograms.
Although the GD is matched, the full bandwidth cannot be supported, resulting in a reduced bandwidth and efficiency.
In contrast, the spectrogram in Fig. \ref{fig:spectrograms_idealized_7fs}(b) the case with matched GDD is shown.
Under ideal conditions, even for a \SI{7}{fs} input pulse, the resulting efficiency can reach up to $\SI{80}{}\%$ with a resulting bandwidth of the SHG pulse of \SI{5.7}{fs} FTL.
The capabilities of matching first- and second order phases while supporting the necessary bandwidth depend on the available mirror coatings.

\section{Conclusion}
In this work we experimentally demonstrate efficient second-harmonic generation of broadband pulses in a multi-pass cell. 
We show that by using the phase-tuning capabilities of MPCs, the SHG-MPC can increase the efficiency of the SHG process for broadband pulses.
In the experiment, we use a thin free-standing BBO crystal placed in the center of a symmetric MPC consisting of two mirrors with GD-compensating optical coatings inside a vacuum chamber.
By optimizing two key parameters, the pressure of the gas in the chamber and the critical angle $\theta$ of the crystal, we demonstrated conversion of $\SI{50}{\micro J}$, \SI{1030}{nm}, \SI{63}{fs} (\SI{56}{fs} FTL) to the SH with \SI{31.5}{fs} FTL with an efficiency of 72\%, exhibiting excellent beam quality.
In a second experiment, we demonstrate SHG for shorter pulses, where we use $\SI{20}{\micro J}$, \SI{1030}{nm}, \SI{15.4}{fs} (\SI{14.6}{fs} FTL) fundamental pulses, converting them into \SI{10}{fs} FTL pulses with an efficiency of 47\% (44\% after compression), which we compress down to \SI{13.8}{fs} at an $M^2=1.18$.
Comparing the bandwidth and resulting efficiency to existing works, we show in Fig.~\ref{fig:overview} that our results outperform common SHG techniques, pushing current efficiency-bandwidth records. The efficiency and bandwidth limits of our method are mainly defined by the achievable mirror dispersion characteristics.  
This combination of high conversion efficiency and broad bandwidth is particularly relevant for applications requiring energetic few-cycle visible or near-visible pulses, such as attosecond science, where efficient second-harmonic generation can help access shorter driving wavelengths and thereby increase the efficiency of high-harmonic generation \cite{Comby2020}.
Our results further indicate that MPCs open new degrees of freedom to tailor second-order nonlinear processes, including not only SHG but also, for example, sum-frequency generation and optical-parametric amplification \cite{Naegele2025, Rajhans2026}, enabling new routes to push the parameter limits of these processes.

\subsection*{Funding \& Acknowledgments}
We acknowledge DESY (Hamburg, Germany), a member of the Helmholtz Association HGF, for support and the provision of experimental facilities. We further acknowledge support from the Helmholtz-Lund International Graduate School (HELIOS) (HIRS-0018) as well as from the Deutsche Forschungsgemeinschaft (DFG, German Research Foundation) – Project 545612524. V.W. acknowledges funding from the German Research Foundation (DFG) - project ID 545611997. We thank Robin Mevert and Prof. Dr. Uwe Morgner, both members of Leibniz Universität Hannover (LUH) for providing critical optical components for the experiments.

\subsection*{Disclosures}
The authors declare no conflicts of interest.

\subsection*{Data Availability Statement}
Data underlying the results presented in this paper are not publicly available but may be obtained from the authors upon reasonable request.
Except where otherwise noted, all figures and images in this document are the original work of the authors and are licensed under a \href{https://creativecommons.org}{CC-BY 4.0 License}.


\clearpage
\onecolumn

\setcounter{section}{0}
\setcounter{subsection}{0}
\setcounter{equation}{0}
\setcounter{figure}{0}
\setcounter{table}{0}

\renewcommand{\thesection}{S\arabic{section}}
\renewcommand{\thesubsection}{\thesection.\arabic{subsection}}

\renewcommand{\theequation}{S\arabic{equation}}
\renewcommand{\thefigure}{S\arabic{figure}}
\renewcommand{\thetable}{S\arabic{table}}

\renewcommand{\theHsection}{S\arabic{section}}
\renewcommand{\theHsubsection}{S\arabic{section}.\arabic{subsection}}
\renewcommand{\theHequation}{S\arabic{equation}}
\renewcommand{\theHfigure}{S\arabic{figure}}
\renewcommand{\theHtable}{S\arabic{table}}

\section*{Supplementary Material}

\section{SHG-MPC simulation model}
\label{sec:supp_3D_model}

\subsection{(3+1)D model}
The simulation model used in this work is based on the Unidirectional Pulse Propagation Equation (UPPE) \cite{Couairon2011}
\begin{align}
    \partial_{z}\,E(k_x,k_y,\omega,z) = i K_z(k_x,k_y,\omega) \,E(k_x,k_y,\omega,z) + iQ(k_x,k_y,\omega)\frac{P(k_x,k_y,\omega,z)}{2\varepsilon_0} \ ,
    \label{eq:UPPE}
\end{align}
where $K_z(k_x,k_y,\omega) = \sqrt{k^2(k_x,k_y,\omega) - k_x^2 - k_y^2}$ and $k(k_x,k_y,\omega) = \frac{\omega}{c_0}n(k_x,k_y,\omega)$, with the spatial frequencies $k_x$ and $k_y$, the frequency $\omega$ and the permittivity of free space $\varepsilon_0 \approx \SI{8.854187819e-12}{F/m}$.
The propagation equation is solved in a non-field resolved manner, where the rapidly oscillating center frequency terms $\bar\omega$ are multiplied out by $\tilde E(\bar\omega + \omega) = E(\omega)\text{e}^{i\bar\omega t}$.
The wavenumber-dependent refractive index is calculated self-consistently according to reference \cite{Lang2013}, giving rise to spatial propagation effects in non-isotropic media, for example and most notably the spatial walk-off of the second harmonic in the crystal \cite{Lang2013}.
The nonlinear propagation coupling is defined as $Q(k_x,k_y,\omega) = \frac{\omega^2}{c^2 K_z(k_x,k_y,\omega)}$, where we make an isotropic approximation $K_z(k_x,k_y,\omega) \approx k(\omega)$ in the nonlinear part only.
The nonlinear polarization $P_{\text{o/e}}(x,y,t,z)$ for ordinary (o) and extraordinary (e) field is solved in the time-domain and takes into account second-order $\chi^{(2)}$ effects as well as self-phase modulation (SPM).
The nonlinear polarization for the second- and third-order nonlinearities is defined as \cite{Lang2013}:
\begin{align}
    P_{\text{o/e}}(x,y,t,z) = P_{\text{o/e}}^{(2)}(x,y,t,z) + P_{\text{o/e}}^{(3)}(x,y,t,z)
\end{align}
\begin{align}
    P_{\text{o}}^{(2)}
    &= \frac{\kappa_o}{4}\,\Big(
        \overbrace{2 E_e E_o^{*}}^{\text{SHG process}} + \ 
        E_e E_e + \ 
        E_e E_e^{*} + \  
        E_o E_e^{*} + \ 
        E_e E_o
    \Big) \\[3pt]
    P_{\text{e}}^{(2)}
    &= \frac{\kappa_e}{4}\,\Big(
        \overbrace{E_o E_o}^{\text{SHG process}} + \ 
        2 E_o E_e^{*} + \ 
        E_e E_o^{*} + \ 
        E_e E_o + \ 
        E_o E_o^{*}
    \Big) \ ,
\end{align}
where we indicate the nonlinear processes that are responsible for the SHG generation in BBO with type I phase-matching.
The pre-factors (coupling parameters) are $\kappa_{o/e} = 2\varepsilon_0\frac{\omega d_{\text{eff}}}{n_{o/e}(\omega)}$, with the $d_{\text{eff}}$ as the medium and its orientation dependent second-order response \cite{Lang2013}.
The additional factor of $1/4$ in the above equations arises from the convention used to calculate the intensity from the electric field $I = \frac{1}{2}nc_0\varepsilon_0|E|^2$, as opposed to $I = 2nc_0\varepsilon_0|E|^2$ used in reference \cite{Lang2013} and by other authors \cite{Boyd2020}. For an explanation of these convention differences, we refer to the preface of reference \cite{Sutherland2003}.
\newline
Finally, the third-order polarization (SPM) is defined as
\begin{align}
    P^{(3)}_{\text{o/e}}(x,y,t,z) = \gamma_{o/e} \, |E_{o/e}(x,y,t,z)|^2 \, E_{o/e}(x,y,t,z) \ ,
\end{align}
with $\gamma_{o/e} = \frac{1}{2}\varepsilon_0^2 c_0 n_{o/e}^2 n_{2, o/e}$.
\noindent

We use an in-house developed code for the numerical evaluation of Equation (\ref{eq:UPPE}), which is solved using a Fourier split-step method. The linear part is solved in reciprocal space and frequency domain and the nonlinear part is solved in real space and time domain using the Runge-Kutta 4 (RK4) method.

\subsection{(1+1)D model}
\label{sec:shg_mpc_numerical_model_1D}

In addition to the (3+1)D model including all spatio-temporal effects, we use a (1+1)D model, where the spatial domain is modeled using the Gaussian ABCD matrix formalism, additionally taking into account Kerr lensing \cite{Hanna2020}.
This hybrid approach enables fast numerical calculations of MPC post-compresison as well as the SHG-MPC by summarizing the two spatial variables $x$ and $y$ into a simple ABCD-matrix based calculation, while capturing the same nonlinear effects of the temporal and spatial domain as the (3+1)D model.
The temporal and spatial domain is modeled by equation \ref{eq:UPPE}, but without the spatial coordinates
\begin{align}
    \partial_{z}\,E(\omega,z) = i k_z(\omega) \,E(\omega,z) + iQ(\omega)\frac{P(\omega,z)}{2\varepsilon_0} \ ,
    \label{eq:UPPE_1D}
\end{align}
where $k_z = \frac{\omega}{c_0}n(\omega)$ and $Q(\omega) = \frac{\omega^2}{c_0^2 k_z}$. Again, nonlinear polarization $P_{\text{o/e}}(t,z)$ for ordinary (o) and extraordinary (e) field is as usually solved in the time-domain and takes into account second-order $\chi^{(2)}$ effects as well as self-phase modulation (SPM).
The nonlinear polarization for the second- and third-order nonlinearities is defined as \cite{Lang2013}:
\begin{align}
    P_{\text{o/e}}(t,z) = P_{\text{o/e}}^{(2)}(t,z) + P_{\text{o/e}}^{(3)}(t,z)
\end{align}
\begin{align}
    P_{\text{o}}^{(2)}
    &= \frac{\kappa_o}{4}\,\Big(
        \overbrace{2 E_e E_o^{*}}^{\text{SHG process}} + \ 
        E_e E_e + \ 
        E_e E_e^{*} + \  
        E_o E_e^{*} + \ 
        E_e E_o
    \Big) \\[3pt]
    P_{\text{e}}^{(2)}
    &= \frac{\kappa_e}{4}\,\Big(
        \overbrace{E_o E_o}^{\text{SHG process}} + \ 
        2 E_o E_e^{*} + \ 
        E_e E_o^{*} + \ 
        E_e E_o + \ 
        E_o E_o^{*}
    \Big) \ ,
\end{align}
where we indicate the nonlinear processes that are responsible for the SHG generation in BBO with type I phase-matching.
The pre-factors (coupling parameters) are again $\kappa_{o/e} = 2\varepsilon_0\frac{\omega d_{\text{eff}}}{n_{o/e}(\omega)}$, with the $d_{\text{eff}}$ as the medium and its orientation dependent second-order response \cite{Lang2013}.
SPM is modeled using 
\begin{align}
    P^{(3)}_{\text{o/e}}(t,z) = \gamma_{o/e} \, |E_{o/e}(t,z)|^2 \, E_{o/e}(t,z) \ ,
\end{align}
with $\gamma_{o/e} = \frac{1}{2}\varepsilon_0^2 c_0 n_{o/e}^2 n_{2, o/e}$.
\newline

\noindent
Finally, the spatial effects are modeled using the ABCD transfer matrices for translation $M_L = \big(\begin{smallmatrix}1 & L/n\\ 0 & 1\end{smallmatrix}\big)$ and focusing $M_f = \big(\begin{smallmatrix}1 & 0\\ -1/f & 1\end{smallmatrix}\big)$.
Kerr lensing is accounted for by calculating the Kerr-focal length $f_{\text{Kerr}}$ as \cite{Hanna2020}:
\begin{align}
    f_{\text{Kerr}} = 0.47 \frac{\pi w^4}{P_{\text{peak}}n_2 dz} \ ,
\end{align}
where $P_{\text{peak}}$ is the peak power.
Kerr lensing is then incorporated by modeling one propagation step of a distance d$z$ as
\begin{align}
    M_{dz} = M_{dz/2} \cdot M_{f_{\text{Kerr}}} \cdot M_{dz/2} \ .
\end{align}

\FloatBarrier
\section{Simulation of experiments}

\begin{figure}[htb!]
    \centering
    \includegraphics[width=1.0\textwidth]{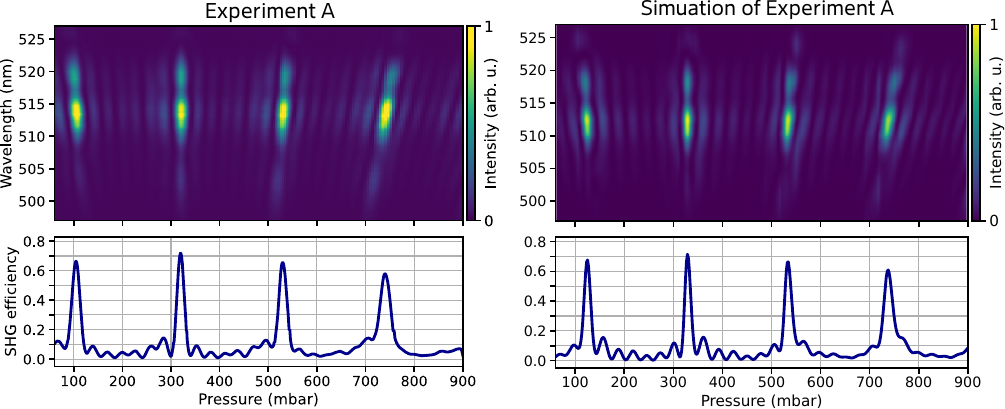}
    \caption{Experimental (left) and simulated (right)  spectrograms for experiment A (section 3A in the main manuscript) using Krypton as the ambient gas for \SI{63}{fs}, \SI{50}{\micro J} fundamental pulses. The data for the input pulse is retrieved from FROG measurements.}
    \label{fig:exp_sim_65fs_kr}
\end{figure}

\begin{figure}[htb!]
    \centering
    \includegraphics[width=1.0\textwidth]{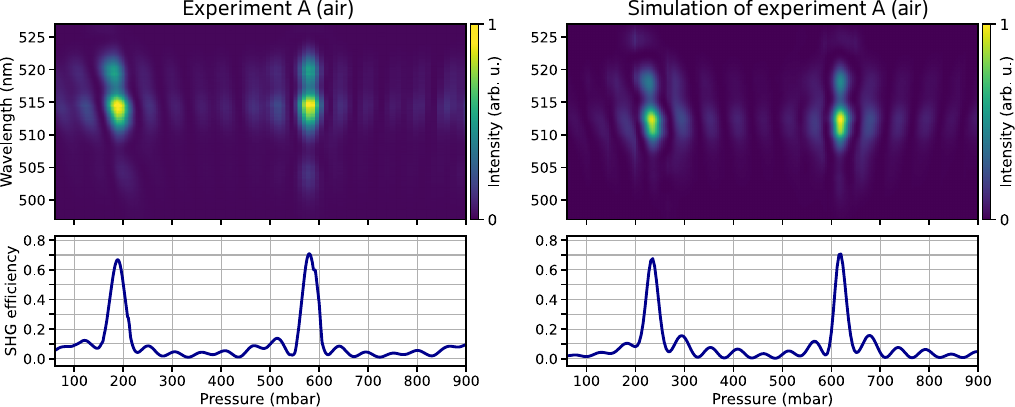}
    \caption{Experimental (left) and simulated (right)  spectrograms for experiment A, but with air as the ambient gas (not shown in the main manuscript) for \SI{63}{fs}, \SI{50}{\micro J} fundamental pulses as the input. The data for the input pulse is retrieved from FROG measurements.}
    \label{fig:exp_sim_65fs_air}
\end{figure}

\begin{figure}[htb!]
    \centering
    \includegraphics[width=1.0\textwidth]{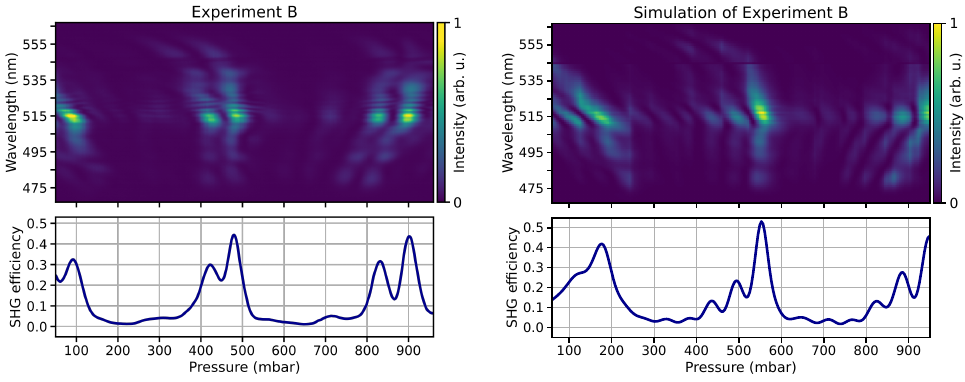}
    \caption{Experimental (left) and simulated (right)  spectrograms for experiment B using air as the ambient gas for \SI{15.4}{fs}, \SI{20}{\micro J} input fundamental pulses. The data for the input pulse is retrieved from FROG measurements.}
    \label{fig:exp_sim_17fs}
\end{figure}

\begin{figure}[htb!]
    \centering
    \includegraphics[width=0.7\textwidth]{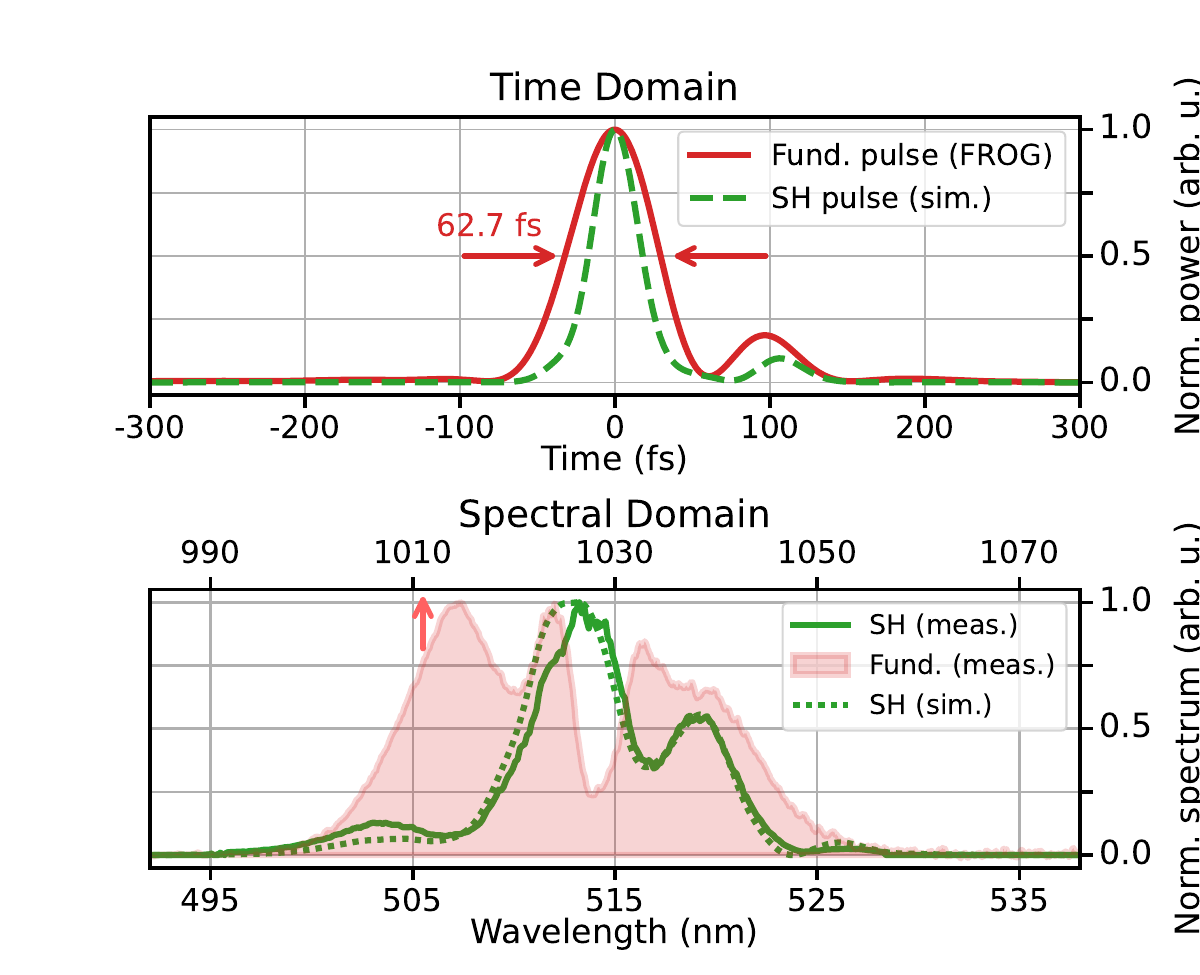}
    \caption{Fundamental and SH pulses from experiment and simulation (Experiment A). The upper plot shows the time domain pulses, the measured fundamental pulse and the compressed simulated SH pulse. In this experiment we did not compress the SH pulses, thus we do not show a compressed measured SH pulse. The lower plot shows the measured input spectrum (red), the measured SH spectrum (green) and the simulated SH spectrum (green, dotted).}
    \label{fig:pulses_exp_sim_65fs_air}
\end{figure}

\begin{figure}[htb!]
    \centering
    \includegraphics[width=0.7\textwidth]{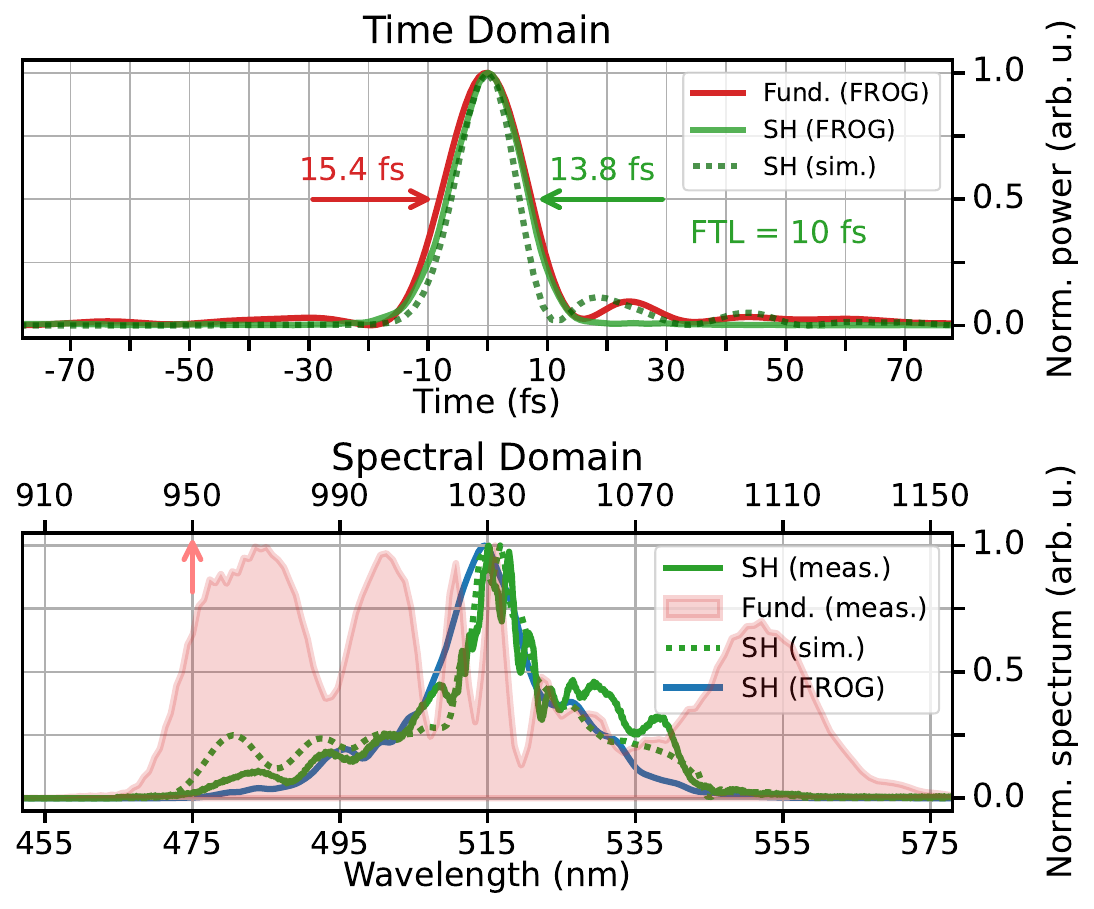}
    \caption{Fundamental and SH pulses from experiment and simulation (Experiment B). The upper plot shows the time domain pulses with the measured fundamental pulse (red), the measured SH pulse (green) and the compressed simulated SH pulse (green, dotted). The lower plot shows the measured input spectrum (red), the measured SH spectrum (green) and the simulated SH spectrum (green, dotted).}
    \label{fig:pulses_exp_sim_17fs_air}
\end{figure}

\begin{figure}[htb!]
    \centering
    \includegraphics[width=1.0\textwidth]{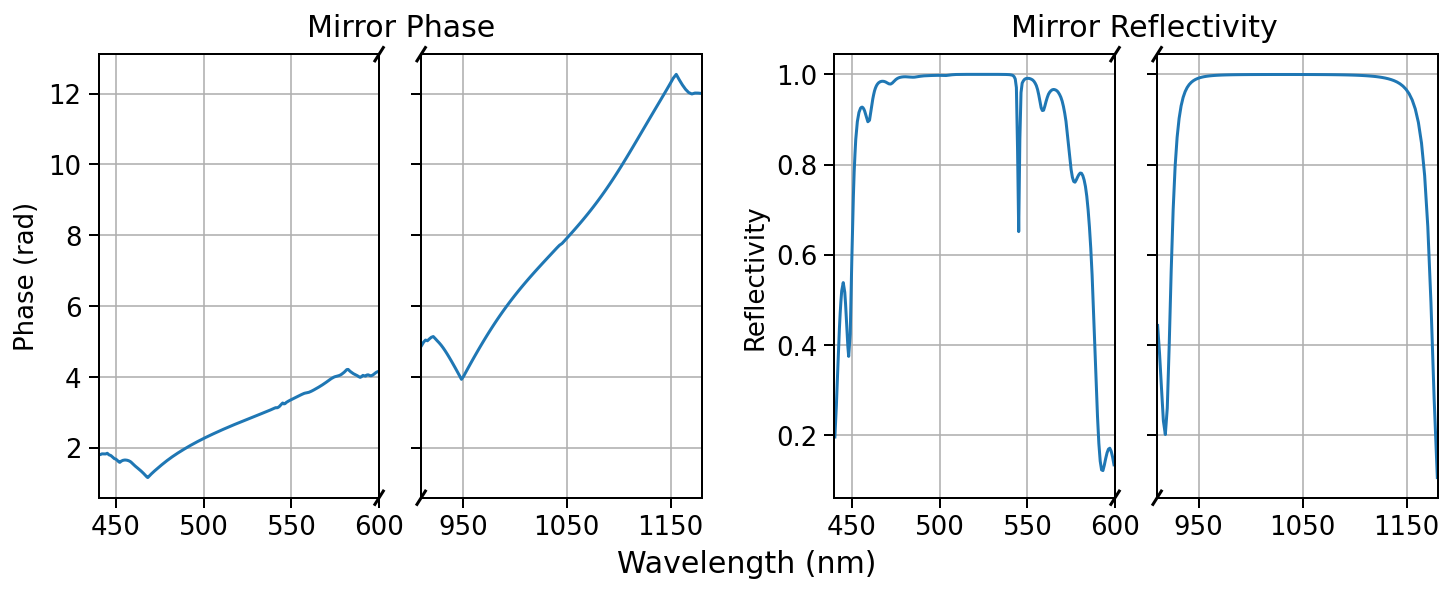}
    \caption{Mirror data for the mirrors used in the experiments and simulations, as provided by the manufacturer LAYERTEC GmbH.}
    \label{fig:mirror_data}
\end{figure}

In this section, we show a collection of numerical results modeling our experimental results. All simulations use the same MPC geometry, consisting of two mirrors with a radius of curvature (ROC) of $R=\SI{1}{m}$, $N=4$ round-trips, and a configuration parameter $k=1$, resulting in an MPC length of $L_{\text{MPC}} = \SI{29.5}{cm}$. 
For the simulations, we use the (3+1)D model described in section \ref{sec:supp_3D_model}.
\newline
\newline
\noindent
Figures \ref{fig:exp_sim_65fs_kr} and \ref{fig:exp_sim_65fs_air} present a comparison between experiment and simulation for experiment A considering a BBO crystal with a thickness of \SI{78}{\micro m}. 
In both datasets, the input pulses have a central wavelength of \SI{1030}{nm}, a duration of \SI{63}{fs}, and an energy of \SI{50}{\micro J}. 
These pulses correspond to retrieved pulse profiles obtained from FROG measurements (Fig. \ref{fig:pulses_exp_sim_65fs_air}), which are used as input for the simulations.
The simulations additionally include the wavelength-dependent phase and reflectivity characteristics of the mirrors, as provided by LAYERTECH GmbH (Fig. \ref{fig:mirror_data}). 
The BBO crystal angle relative to the optical axis is set to $\theta=\SI{24}{\degree}$, and the gas pressure is varied between approximately \SI{60}{mbar} and \SI{900}{mbar}.
In both cases, excellent agreement between simulation and experiment is observed in the spectrograms as well as in the spectra. 
Furthermore, the measured second-harmonic (SH) spectrum supports a Fourier transform-limited (FTL) pulse duration of \SI{31.5}{fs}, which is in exact agreement with the value obtained from the simulated SH spectrum.
We furthermore incorporate an estimated anti-reflective (AR) coating loss of 1\% for each field, fundamental and SH, which yields a maximum efficiency of 71\% for both gases, which is very close to the measured maximum efficiency of 72\%.
\newline
\newline
\noindent
Figure \ref{fig:exp_sim_17fs} shows the experimental and simulated spectrograms for experiment B (see Section 3B of the main manuscript), using air as the ambient gas medium and fundamental input pulses at \SI{1030}{nm} with a duration of \SI{15.4}{fs} and an energy of \SI{20}{\micro J}. 
These input pulses were characterized using FROG measurements (Fig. \ref{fig:pulses_exp_sim_17fs_air}) and used as input for the simulations. 
In this simulation, the BBO crystal thickness was increased to \SI{95}{\micro m} and the $\theta$ angle is set to $23.2^{\circ}$.
Furthermore, the measured and simulated SH spectra show good agreement, with both supporting a FTL pulse duration of \SI{10}{fs}.
The secondary maxima are reproduced as well, when the angular variation in the critical $\theta$ direction resulting from the MPC geometry is accounted for.
As discussed in the main manuscript, the spectra retrieved from the FROG measurements exhibit a reduced bandwidth. 
This narrowing can be attributed to the limited phase-matching bandwidth of the BBO crystal used in the FROG setup (Fig. \ref{fig:pulses_exp_sim_17fs_air}).
In this simulation we apply an estimated AR-coating loss of 2\% for each field, which results in a maximum conversion efficiency of about 53\%, which is reasonably close to the measured efficiency of 47\%.


\FloatBarrier
\section{Second-harmonic beam quality measurements}

\begin{figure}[htb!]
    \centering
    \includegraphics[width=0.8\textwidth]{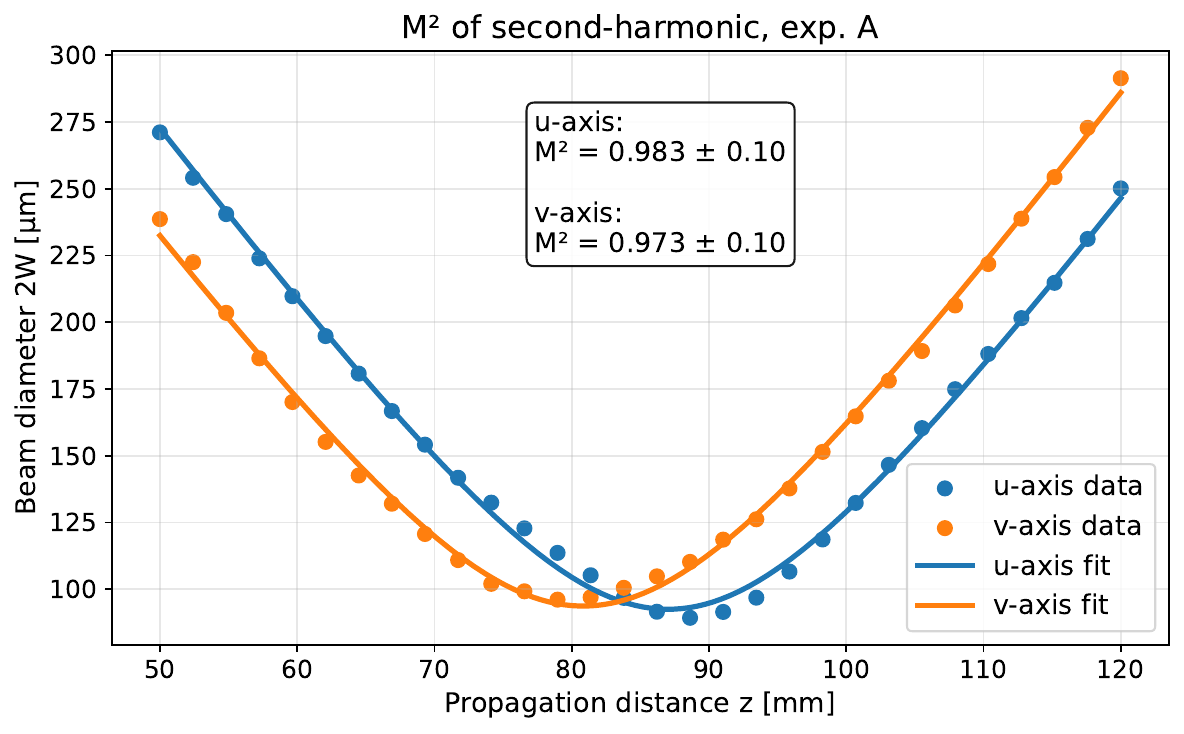}
    \caption{$M^2$ measurement of the output Ssecond-harmonic beam in experiment A (long pulse). }
    \label{fig:M2_expA}
\end{figure}

\begin{figure}[htb!]
    \centering
    \includegraphics[width=0.8\textwidth]{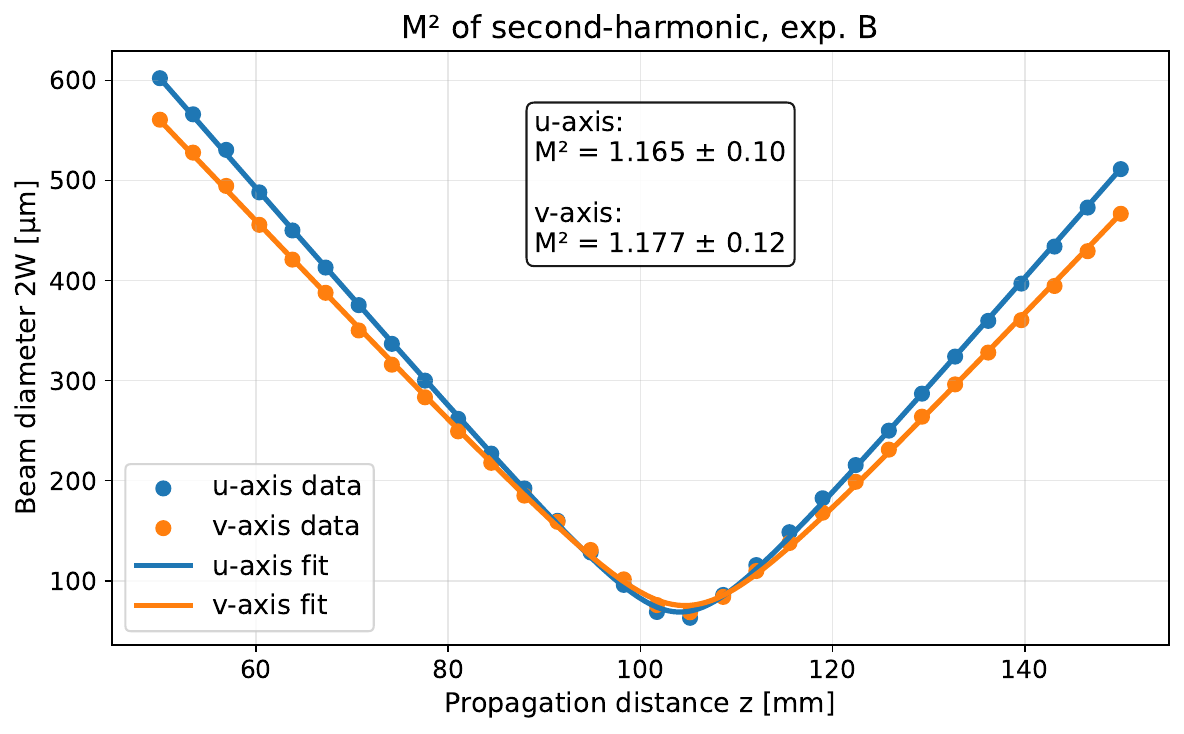}
    \caption{$M^2$ measurements of the output second-harmonic beam in experiment B (short pulse).}
    \label{fig:M2_expB}
\end{figure}

\newpage
\bibliography{shg_mpc_bibliography_titles_cleaned}

\end{document}